\documentclass[12pt]{iopart}

\newcommand{\be}   {\begin{equation}}
\newcommand{\ee}   {\end{equation}}
\newcommand{\ba}   {\begin{eqnarray}}
\newcommand{\ea}   {\end{eqnarray}}

\begin{document}

\title{Open orbits and the semiclassical dwell time}     

\author{C. H. Lewenkopf\dag\ and R. O. Vallejos\ddag\ }

\address{\dag\ Instituto de F\'{\i}sica, 
	         Universidade do Estado do Rio de Janeiro,\\ 
	         R. S\~ao Francisco Xavier 524, 
	         20550-900 Rio de Janeiro, Brazil}

\address{\ddag\ Centro Brasileiro de Pesquisas F\'{\i}sicas, \\
                R. Dr. Xavier Sigaud 150, 
                22290-180 Rio de Janeiro, Brazil}

\date{\today}

%%%%%%%%%%%%%%%%%%%%%%%%%%%%%%%%%%%%%%%%%%%%%%%%%%%%%%%%%%%%%%%%%%%%%%%%%%%%
\begin{abstract}
The Wigner delay time is addressed semiclassically using the 
Miller's $S$-matrix expressed in terms of open orbits.
This leads to a very appealing expression, in terms of classical paths, 
for the energy averaged Wigner time delay in chaotic scattering.
The same approach also puts in evidence the semiclassical
incapability to correctly assess the time delay higher moments.
This limitation suggests that the use of the semiclassical 
approximation to quantify fluctuations in scattering phenomena, 
like in mesoscopic physics, has to be considered with great caution.
\end{abstract}
%%%%%%%%%%%%%%%%%%%%%%%%%%%%%%%%%%%%%%%%%%%%%%%%%%%%%%%%%%%%%%%%%%%%%%%%%%%%

%\pacs{05.45.Mt}
% 05.45.Mt  Semiclassical chaos ("quantum chaos")

\maketitle

%------------------------------------------------------------------------- 

What is the time a quantum particle spends to traverse a scattering
region?
Reported superluminal wave propagation \cite{Nature} and
its relation to the fascinating and controversial problem of the 
``tunneling time" \cite{Buttiker82JPA,Buttiker83JPA,Nussenzveig00JPA} 
brought renewed interest to the quantum (or wave) dwell time 
question in general. 
The status of this matter is very nicely discussed by a recent 
comprehensive and pedagogical review \cite{Carvalho02JPA}.

The present study is devoted to examine the delay time of a particle 
scattered by a chaotic potential \cite{Smilansky91JPA,Lewenkopf91JPA}
in the absence of tunneling barriers.
Here the semiclassical approximation is employed to take a fresh look 
over the Wigner time delay \cite{Wigner55JPA}.
Starting directly from the Miller semiclassical $S$ matrix, a quite 
appealing expression for the average dwell time is obtained and
some quite unexpected limitations to the semiclassical approximation 
are clearly revealed.

The Wigner-Smith time delay matrix \cite{Wigner55JPA,Smith60JPA} 
is defined as
\be
\label{eq:Q_ab}
Q_{ab}(E)=-i\hbar \sum_{c=1}^N S_{ac}^\ast \frac{\partial S_{cb}}
{\partial E}  ~,
\label{eq:partial_time}
\ee
where the scattering matrix $S$, that encodes all accessible information 
about the scattering process, is taken at the energy $E$. The sum in
(\ref{eq:Q_ab}) runs over all $N$ open asymptotic scattering channels.
The Wigner time delay $\tau_W(E)$ is then given as
\be
\tau_W(E)= \frac{1}{N}\,\mbox{tr}\, Q (E)\;.
\label{eq:time}
\ee

In scattering processes large enhancements of $\tau_W(E)$, or long dwell 
times, correspond to narrow isolated resonances. In these situations, the 
Wigner time delay gives access  to properties of individual quasi-bound 
states. In the opposite and also relevant case of overlapping resonances, 
information about individual states is lost. The fluctuations of 
$\tau_W(E)$ result from the coherent interference of many resonances. 
Here the scattering process is best characterized by suitable statistical
measures of $\tau_W(E)$ averaged over energy windows containing many
resonances. 
In chaotic systems $\tau_W$ displays universal fluctuations that can
be described by the theory of random matrices. A very detailed discussion
on various aspects on the statistical approach for the Wigner time
delay can be found, for instance, in \cite{Fyodorov97JPA}.

The semiclassical literature on the Wigner time delay is invariantly 
based on its relation to the level density \cite{Eckhardt93JPA,Vallejos98JPA}, 
as envisaged long time ago by Friedel \cite{Friedel52JPA,Balian74JPA}.
Hence, the resulting semiclassical $\tau_W$ depends solely on the properties 
of the periodic orbits trapped in the scattering region.
The semiclassical calculated Wigner time fluctuations coincide with the 
random matrix results \cite{Lehmann95aJPA,Lehmann95bJPA} for $N \gg 1$, 
as they should.

It is also desirable to cast the semiclassical $\tau_W$ in terms of 
open classical orbits that spend a finite time in the scattering region, 
closer to the spirit of a scattering problem.
This is the task pursued here.
Equation (\ref{eq:Q_ab}) is directly evaluated using the Miller's semiclassical 
$S$-matrix \cite{Miller75JPA}, namely
\be
\label{eq:Miller}
    \widetilde{S}_{ab}(E) = \sum_{\mu(a \leftarrow b)} \sqrt{p_\mu(E)} \, 
         e^{i\sigma_\mu(E)/\hbar} \;,
\ee
where the classical trajectories that start at channel $b$ and end at 
channel $a$ are labelled by $\mu(a \leftarrow b)$. Accordingly, $\sigma_\mu$
is the reduced action (with a Maslov phase included) and $p_\mu$ is the 
classical transition probability for going from $b$ to $a$ following the 
path $\mu$ \cite{Smilansky91JPA}.
Throughout this paper the wide tilde indicates quantities obtained by means
of the semiclassical approximation.
In the derivation of (\ref{eq:Miller}) the absence of tunneling barriers 
between the scattering and the asymptotic regions is implicit. Furthermore,
the number of open channels $N$ must fulfill $N \gg 1$. 

For the semiclassical Wigner time delay, energy variations in 
classical transition probabilities $p_\mu$ are negligible as compared 
to those in the actions $\sigma_\mu$, since the latter are measured in 
units of $\hbar$. Hence,
\be
\widetilde{\tau}_W
  =  \frac{1}{N}\sum_{\mu,\nu} t_\mu 
     \sqrt{p_\mu p_\nu} \,  e^{i(\sigma_\mu - \sigma_\nu)/\hbar} \;,
\ee
where $t_\mu = \partial \sigma_\mu /\partial E$ is the classical time 
the particle spends to go from $b$ to $a$ through the path $\mu$. Here 
the sums are unrestricted and run over {\sl all} classical trajectories 
that enter and leave the scattering region.
It is worth stressing that, due to the semiclassical approximation,
$\widetilde{\tau}_W$ is not manifestly real for any given value of 
energy $E$.
In analogy to the unitarity deficit of the semiclassical $S$-matrix 
\cite{Vallejos01JPA}, it can be shown that the imaginary part of 
$\widetilde{\tau}_W$ is of subleading order in powers of $1/N$.
This spurious imaginary part can be easily eliminated by using, 
for instance,
\begin{equation}
\label{eq:Q_absym}
Q^{\rm sym}_{ab}(E) = -i \hbar \sum_c \,\frac{d}{d\varepsilon}\!\left.\left[
              S_{ac}(E + \frac{\varepsilon}{2})
              S^*_{ac}(E - \frac{\varepsilon}{2}) \right]\right|_{\varepsilon=0} 
\end{equation}
instead of $Q_{ab}$ defined by (\ref{eq:Q_ab}). 
Restricting the analysis to the leading order term in powers of $1/N$, it is 
possible to insist with $Q_{ab}$. In addition to the simplicity, this strategy 
has the merit of exposing some of the semiclassical limitations.

The statistical analysis of $\widetilde{\tau}_W$ and its higher moments unveils 
system specific, as well as universal features \cite{Carvalho02JPA}.
Let us start discussing the energy averaged $\widetilde{\tau}_W$, namely
\be
\langle \widetilde{\tau}_W \rangle  
  =  \frac{1}{N}\sum_{\mu,\nu} 
     \left\langle t_\mu 
     \sqrt{p_\mu p_\nu} \, 
      e^{i(\sigma_\mu - \sigma_\nu)/\hbar} 
     \right\rangle \; .
\ee
Here $\langle \cdots \rangle$ indicates an energy average taken over an 
energy window $\Delta E$ where the classical dynamics presents little
changes, nonetheless comprising many resonances. 
To compute $\langle \widetilde{\tau}_W \rangle$ it is justified to neglect 
the energy dependence of the probabilities $p_\mu$ and use the diagonal 
approximation.
The latter says that, in general, different orbits of a chaotic system are 
uncorrelated, and holds for trajectories with dwell times shorter than 
the Heisenberg time $\tau_{\rm H}\equiv h/\Delta$. 
Here $\Delta$ is the mean resonance spacing.
Fortunately, without barriers, trajectories with $t$ exceeding $\tau_H$ 
are statistically negligible in the semiclassical regime of $N\gg 1$ 
\cite{Vallejos98JPA}.
In the absence of system specific symmetries the diagonal 
approximation reads $\left\langle \exp[i(\sigma_\mu - \sigma_\nu)/\hbar] 
\right\rangle = \delta_{\mu \nu}$,
yielding 
\be
\label{eq:simple}
 \langle \widetilde{\tau}_W \rangle =
 \frac{1}{N}\sum_{\mu,\nu} t_\mu \sqrt{p_\mu p_\nu} \, \delta_{\mu \nu}
 = \frac{1}{N}\sum_{\mu} p_\mu t_\mu \equiv \tau                      \; .
\ee
This remarkable equation expresses the multichannel energy averaged 
Wigner time delay purely in terms of classical quantities: $\tau$ 
is the classical scattering region escape time obtained by averaging 
over all trajectories weighted by their transition probabilities. 
The energy average eliminates the quantum interference terms
in the leading order. 
Equation (\ref{eq:simple}) holds for perfect transmission, that is
guaranteed in the semiclassical limit by the absence of tunneling
barriers.
The diagonal approximation is customarily justified only for chaotic systems.
Hence, chaos is a key element to obtain (\ref{eq:simple}). 
The validity of the diagonal approximation for other kinds of dynamics is
unclear. 
Thus, despite of the simplicity and appeal of (\ref{eq:simple}), 
a rigorous derivation of an expression of the same kind for 
integrable systems is still lacking.

This (semi)classical result raises an important question:
Is $\langle \widetilde{\tau}_W \rangle = h/(N\Delta)$ consistent with the 
well-known exact identity $\langle \tau_W(E) \rangle = h/(N\Delta)$, 
that holds irrespective of the underlying classical dynamics 
\cite{Lewenkopf91JPA}? In other words, is it possible to prove
$\tau = h/(N\Delta)$ using just geometric arguments? 

The equivalence between both relations can be shown for the special case 
of chaotic billiards connected to the scattering region by wave guides.
The proof is straightforward.
The average time between bounces $\tau_{\rm b}$ for an ergodic billiard 
is 
\be
\label{eq:tbounce}
\tau_{\rm b}=\frac{\pi A}{Pv} \quad \mbox{for}\,\,\,d=2
\qquad \mbox{and} \qquad 
\tau_{\rm b}=\frac{4 V}{Av}   \quad  \mbox{for}\,\,\,d=3
\ee
where $v=\sqrt{2E/m}$ is the particle velocity. In two dimensions ($d=2$), 
$A/P$ is the billiard surface to perimeter ratio. Likewise, $V/A$ is the 
volume to surface ratio for three-dimensional ($d=3$) cavities.
The above relations are {\sl exact}. 
As discussed in \cite{Chernov97JPA}, their rigorous derivation is long known 
by mathematicians.
Simple physical heuristic arguments can also used to obtained 
(\ref{eq:tbounce}), as shown in \cite{Jarzynski93JPA}.

For ergodic billiards the relation between the scape time $\tau$ and the 
average bounce time $\tau_{\rm b}$ is
\be
\label{eq:tscape}
\tau = \frac{P}{W} \,\tau_{\rm b}\quad \mbox{for}\,\,\,d=2
\qquad \mbox{and} \qquad 
\tau = \frac{A}{S} \,\tau_{\rm b}\quad \mbox{for}\,\,\,d=3
\ee
where $W$ and $S$ are respectively the wave guide width in $d=2$ and 
its cross section area in $d=3$. 
The Weyl formula can be used twice to relate the geometry of the billiard
to the mean resonance spacing $\Delta$ and the 
number of channels in the wave guides $N$, yielding
\be
\tau = \frac{1}{N} \frac{h}{\Delta}
\ee
for both two and three dimensional systems, as previously announced.

Let us now use the ``open orbits" semiclassical approximation  
to calculate the Wigner time delay autocorrelation function, 
namely
\be
C(\epsilon)= \langle \tau_W(E+\epsilon) \tau_W(E) \rangle_E  - 
\langle \tau_W(E) \rangle^2~.
\ee
More explicitly, we compute
\be
\label{eq:Cexplicit}
\fl
\widetilde{C}(\epsilon)=
\frac{1}{N^2}\sum_{{a,b}\atop{c,d}}\sum_{{\mu,\nu(a\leftarrow b)}\atop
{\mu^\prime,\nu^\prime(c\leftarrow d)}} \sqrt{p_\mu p_\nu p_{\mu^\prime} 
p_{\nu^\prime}} t_\mu t_{\mu^\prime} \left\langle e^{\frac{i}{\hbar}
(\sigma_\mu - \sigma_\nu + \sigma_{\mu^\prime} - \sigma_{\nu^\prime})}
\right\rangle e^{\frac{i}{\hbar}(t_\mu - t_\nu) \epsilon} - 
\langle \widetilde{\tau}_W(E) \rangle^2~,
\ee
and compare it with the result obtained using either the stochastic approach 
\cite{Lehmann95aJPA,Lehmann95bJPA}, or
the ``closed orbits" semiclassical theory \cite{Eckhardt93JPA}.
The analysis of the transition probabilities $p_\mu$ gives the answer 
almost immediately.
For chaotic systems the transition probabilities follow the analogue 
of the Hannay-Ozorio de Almeida sum rule for open systems 
\cite{Kadanoff84JPA,Hannay84JPA}
\be
\label{eq:sumrule}
\sum_{t \le t_\mu \le t + \delta t} p_\mu = \frac{1}{N\tau}\,
\mbox{e}^{-t/\tau} \delta t
\ee
where $\sum_{t \le t_\mu \le t + \delta t}\,p_\mu$ is the sum of 
all classical transition probabilities following the trajectories 
$\mu$ belonging to the a small time interval $[t, t+\delta t]$, 
where $\delta t$ is classically small. 
In (\ref{eq:sumrule}) the sum has no restriction on channels. Hence,
it is easy to verify that by replacing the sum by a time integral, 
the classical normalization (flux conservation) condition \cite{Smilansky91JPA} 
\be
\label{eq:classicalnormalization}
\sum_{a=1}^N \sum_{\mu(a \leftarrow b)} p_\mu = 1 \;,
\ee
is fulfilled. Note that (\ref{eq:classicalnormalization}) must hold 
irrespective of the system dynamics. 

The semiclassical Wigner time delay autocorrelation function $\widetilde{C}(\epsilon)$
can be calculated using the diagonal approximation
\be
\label{eq:superdiagonal}
\left\langle e^{\frac{i}{\hbar}
(\sigma_\mu - \sigma_\nu + \sigma_{\mu^\prime} - \sigma_{\nu^\prime})}
\right\rangle = \delta_{\mu \nu       } \delta_{\mu^\prime \nu^\prime} + 
                  \delta_{\mu \nu^\prime} \delta_{\nu        \mu^\prime} 
\ee
and the sum rule (\ref{eq:sumrule}). First, the diagonal approximation contracts 
pairwise the summations over the orbit indices in (\ref{eq:Cexplicit}). 
As a result $\langle \widetilde{\tau}_W(E) \rangle^2$ is cancelled and 
only a double sum over the paths remains to be evaluated. By grouping the 
paths with similar traversal times we then use the sum rule (\ref{eq:sumrule}) 
to transform the summations over orbits into time integrals. 
The later immediately give
\be
\label{eq:naiveC}
\widetilde{C}(\epsilon) = 
\frac{\tau^2}{N^2} \frac{1}{\left[1 + \left(\frac{\epsilon \tau}{\hbar}
\right)^2 \right]^2} \,.
\ee
Since no special attention is payed to time-reversal symmetric paths,
(\ref{eq:naiveC}) represents the semiclassical correlation function for 
broken time-reversal symmetry.
It is noteworthy that although $\widetilde{\tau}_W(E)$ is not purely 
real, the energy average eliminates the imaginary part of $\widetilde{C}
(\epsilon)$.
However, (\ref{eq:naiveC}) is at odds with the random matrix result 
\cite{Lehmann95aJPA,Lehmann95bJPA}, that reads
\be
\label{eq:Cexact}
C(\epsilon) = 
\frac{2\tau^2}{N^2} \frac{1 - \left(\frac{\epsilon \tau}{\hbar}
\right)^2 }
   {\left[1 + \left(\frac{\epsilon \tau}{\hbar}
\right)^2 \right]^2} \,,
\ee
in the limit of $N\gg 1$ and perfect transmission.
The agreement is not reestablished by just starting with a 
Wigner-Smith matrix $Q$ which in the semiclassical limit is
manifestly Hermitian, as for instance $Q^{\rm sym}$ in 
(\ref{eq:Q_absym}). 
Neither the variance, nor the $\epsilon$ dependence
of the Wigner time correlation function defined as 
$\widetilde{\tau}^{\rm sym}_W = (1/N)\tr Q^{\rm sym}$ agrees
with (\ref{eq:Cexact}).

Related problems appear when the semiclassical approximation 
using open orbits is employed to obtain the average transmission 
through a cavity and its autocorrelation functions, quantities of 
central interest in mesoscopic quantum coherent electronic transport. 
In this case the standard semiclassical approach fails because 
the $S$ matrix is not unitary. 
As previously discussed \cite{Vallejos01JPA} this problem is best 
characterized by noticing that
\be
\left\langle \left(\sum_{c=1}^N |\widetilde{S}_{ac}(E)|^2 - 1\right)
\left(\sum_{d=1}^N |\widetilde{S}_{bd}(E)|^2 - 1\right)\right\rangle
= \frac{\delta_{ab}}{N} \neq 0.
\ee
Although admittedly small, the lack of unitarity is of the same order
as typical transmission correlation functions. 
These spurious ``unitarity" fluctuations could perhaps be reduced by 
including higher order corrections in the individual semiclassical 
$S$ matrix elements. 
However, a correct description of the correlations among different matrix 
elements, neglected in the standard semiclassical approach, results to 
be a more practical way to proceed.
In fact, for the specific case of transmission, the 
discrepancies between random matrix theory and the semiclassical 
approximation could be fixed by introducing proper semiclassical sum 
rules that impose unitarity \cite{Vallejos01JPA}. 

The Wigner time delay problem does not seem to have such a simple 
solution.
Here the shortcomings of the semiclassical approach are more 
severe. 
For instance, a direct inspection shows that both the variance and 
the functional dependence of $\langle\widetilde{\tau}_W(E+\epsilon)
\widetilde{\tau}_W(E)\rangle$ and $\langle\widetilde{\tau}_W(E+
\epsilon)\widetilde{\tau}_W^*(E)\rangle$ are very different.
Unfortunately, even (\ref{eq:Q_absym}) which seemed 
very promising to eliminate the semiclassical spurious imaginary 
part feature is of little help: Although $\widetilde{\tau}^{\rm sym}_W(E)$
is manifestly real, $\widetilde{\tau}^{\rm sym}_W(E)
\widetilde{\tau}^{\rm sym}_W(E+\epsilon)$ is not.

In distinction to the approach presented in this study, the ``closed
orbits" semiclassical approximation to the Wigner time delay is very 
successful to describe its fluctuations \cite{Eckhardt93JPA,Vallejos98JPA}.
The reason is simple. The trace over $Q$ allows one to express
the Wigner time delay as a density of states, see for instance 
\cite{Lehmann95aJPA}.
This is a more convenient starting point for the semiclassical 
approximation: the density of states is manifestly real
and since $Q$ was already traced the unitarity problems of
$\widetilde{S}$ do not appear.
As a result one keeps only the contributions of closed orbits trapped 
in the scattering region, loosing all information about the channels.
The small price to pay is that the simple physical interpretation for 
the average $\widetilde{\tau}_W$ is missed. 
On the other hand, since numerous observables of interest in scattering 
problems do not involve traces over all channels, a more general 
solution is desirable.

It is important to mention that for some classes of cross-section correlations,
the relation between random-matrix theory \cite{Fyodorov98JPA} and semiclassical 
results is in very close relation to the above discussion. 
As nicely addressed by Ref.\ \cite{Eckhardt00JPA}, apparent discrepancies
between the two approaches are successfully remedied by avoiding taking 
singular semiclassical limits.
Unfortunately we do not see how to adapt such ideas to our problem: The scattering 
problem considered in \cite{Eckhardt00JPA} allows for the use of a ``closed orbits"-like 
semiclassical approximation that is free of the unitarity problems discussed here. 

In summary, our study shows that the semiclassical $S$ matrix leads to a very 
appealing expression, in terms of classical paths, for the 
energy averaged Wigner time delay in chaotic scattering.
On the other hand, it also puts in clear evidence the semiclassical
incapability to correctly assess the time delay higher moments.
This limitation can be attributed to the spurious imaginary part 
of $\widetilde{\tau}_W$ and to the lack of unitarity of 
the $\widetilde{S}$ matrix.
The unitarity problem is quite severe and not only specific to the 
object studied in this work. 
Hence, before this problem is circumvented, applications of the 
semiclassical approximation to quantify fluctuations in scattering 
phenomena, like in mesoscopic physics, have to be considered with 
great caution.

\ack 
The authors thank CNPq and PRONEX (Brazil) for partial financial support.

%-----------------------------------------------------------------------


\Bibliography{99}
%-----------------------------------------------------------------------

\bibitem{Nature}
   Wang L J, Kuzmich A, Dogaiu A 2000
        {\it Nature} {\bf 406}, 277 \\ 
   Steinberg A M Kwiat P G Chiao R Y 1993
        {\it Phys. Rev. Lett.} {\bf 71} 708 

\bibitem{Buttiker82JPA}
    B\"uttiker M and Landauer R 1982
         {\it Phys. Rev. Lett.} {\bf 49} 1739

\bibitem{Buttiker83JPA}
    B\"uttiker M 1983
         {\it Phys. Rev.} B {\bf 2} 6178

\bibitem{Nussenzveig00JPA}
    Nussenzveig H M 2000
         {\it Phys. Rev.} A {\bf 62} 042107
% average dwell time

\bibitem{Carvalho02JPA}
   de Carvalho C A A and Nussenzveig H M 2002
        {\it Phys. Rep.} {\bf 364} 83

\bibitem{Smilansky91JPA}
   Smilansky U 1991 
        {\it Les Houches 1989 Session LII on Chaos and 
             Quantum Physics} ed  M J Gianonni, A Voros and 
             J Zinn-Justin 
        (Amsterdam: North-Holland) pp 371-441     

\bibitem{Lewenkopf91JPA}
   Lewenkopf C H and Weidenm\"uller H A 1991
         {\it Ann. Phys. (N.Y.)} {\bf 212} 53

\bibitem{Wigner55JPA}
   Wigner E P 1955
        {\it Phys. Rev.} {\bf 98} 145

\bibitem{Smith60JPA}
   Smith F T 1960
        {\it Phys. Rev.} {\bf 118} 349

\bibitem{Fyodorov97JPA}
   Fyodorov Y V and Sommers H J 1997
         {\it J. Math. Phys.} {\bf 38} 1918

\bibitem{Eckhardt93JPA}
   Eckhardt B 1993
         {\it Chaos} {\bf 3} 613

\bibitem{Vallejos98JPA}
  Vallejos R O, Ozorio de Almeida A M and Lewenkopf C H 1998
        {\it J. Phys. A: Math. Gen.} {\bf 31} 4885

\bibitem{Friedel52JPA}
    Friedel J 1952
         {\it Philos. Mag.} {\bf 43} 153

\bibitem{Balian74JPA}
    Balian R and Bloch C 1974
         {\it Ann. Phys. (N.Y.)} {\bf 85} 514

\bibitem{Lehmann95aJPA}
    Lehmann N, Savin D, Sokolov V and Sommers H-J 1995
         {\it Physica} {\bf 86D} 572

\bibitem{Lehmann95bJPA}
    Lehmann N, Savin D, Sokolov V and Sommers H-J 1995
         {\it Nucl. Phys.} {\bf A582} 223

\bibitem{Miller75JPA}
   Miller W H 1975
     {\it Advances in Chemical Physics} ed K. P. Lawley
     (New York: Wiley) Vol. 30 p 77

\bibitem{Vallejos01JPA}
   Vallejos R O and Lewenkopf C H 2001
         {\it J. Phys. A: Math. Gen.} {\bf 34} 2713 

\bibitem{Chernov97JPA}
   Chernov N 1997
        {\it J. Stat. Phys.} {\bf 88} 1

\bibitem{Jarzynski93JPA}
   Jarzynski C 1993 
        {\it Phys. Rev.} E {\bf 48} 4340

\bibitem{Kadanoff84JPA}
   Kadanoff L P and Tang C 1984 
        {\it Proc. Natl. Acad. Sci. USA} {\bf 81} 1276 

\bibitem{Hannay84JPA}
   Hannay J H and Ozorio de Almeida A M 1984 
        {\it J. Phys. A: Math. Gen.} {\bf 17} 3429   

\bibitem{Fyodorov98JPA}
   Fyodorov Y V and Alhassid Y 1998  
        {\it Phys. Rev.} A {\bf 58} R3375 \\
   Alhassid Y and Fyodorov Y V  1998  
        {\it J. Phys. Chem.} {\bf 102} 9577

\bibitem{Eckhardt00JPA}
   Eckhardt B, Fishman S and Varga I 2000
        {\it Phys. Rev.} E {\bf 62} 7867 \\
   Eckhardt B, Varga I and Pollner P 2001  
        {\it Physica} E {\bf 9} 535
      

\endbib
\end{document}